\documentclass[prb,aps,twocolumn,amsmath,amssymb,floatfix,
superscriptaddress]{revtex4-1}
\usepackage[dvips]{graphics}
\usepackage{graphicx}% Include figure files
\usepackage{color}
\usepackage{blindtext}
\definecolor{dred}{rgb}{0.75,0,0}
\usepackage{graphicx}% Include figure files
\usepackage{dcolumn}% Align table columns on decimal point
\usepackage{bm}% bold math
\usepackage{xcolor}
\usepackage{appendix}
\usepackage{hyperref}% add hypertext capabilities
\begin{document}

\preprint{APS/123-QED}

\title{\textcolor{dred}{Engineering topological phase transition and Aharonov-Bohm caging in a flux-staggered lattice}} 

\author{Amrita Mukherjee}
\affiliation{Department of Physics, University of Kalyani, Kalyani,
West Bengal-741 235, India}
\email{amritaphy92@gmail.com}

\author{Atanu Nandy}
\affiliation{Department of Physics, Kulti College, Kulti, Paschim Bardhaman,
West Bengal-713 343, India}
\email{atanunandy1989@gmail.com}

\author{Shreekantha Sil}
\affiliation{Department of Physics, Visva-Bharati, Santiniketan, 
West Bengal-731 235, India}
\email{shreekantha.sil@visva-bharati.ac.in}

\author{Arunava Chakrabarti}
\affiliation{Department of Physics, Presidency University, 86/1 College Street, Kolkata, West Bengal - 700 073, India}
\email{arunava.physics@presiuniv.ac.in}
\date{\today}% It is always \today, today,

\begin{abstract}
A tight binding network of diamond shaped unit cells trapping a staggered magnetic flux distribution is shown to exhibit a topological phase transition under a controlled variation of the flux trapped in a cell. A simple real space decimation technique maps a binary flux staggered network into an equivalent Su-Shrieffer-Heeger (SSH) model. In this way, dealing with a subspace of the full degrees of freedom, we show that a topological phase transition can be initiated by tuning the applied magnetic field that eventually simulates an engineering of the numerical values of the overlap integrals in the paradigmatic SSH model. Thus one can use an external agent, rather than monitoring the intrinsic property of a lattice to control the topological properties. This is advantageous from an experimental point of view. We also provide an in-depth description and analysis of the topologically protected edge states, and discuss how, by tuning the flux from outside one can enhance the spatial extent of the Aharonov-Bohm caging of single particle states for any arbitrary period of staggering. This feature can be useful for the study of transport of quantum information. Our results are exact.
\end{abstract}

\maketitle

%\tableofcontents
\section{Introduction}
The Su-Schrieffer-Heeger (SSH) model, describing spinless fermions on a one dimensional lattice where the electron hopping is staggered,~\cite{ssh,heeger} stands out as a paradigmatic example of a one-dimensional charge fractionalized  and topologically ordered system.~\cite{tknn,prange,wen} Symmetry protected topological order, a hallmark of the topological insulators (TI), has enriched and revolutionised concepts in condensed matter physics by unravelling a novel state of matter where the energy bands are characterized by quantized topological invariants. Needless to say, the advent of this very special concept of topological order and an exotic quantum phase in condensed matter systems have spurred immense research activity, both in theory and in experiments, merging the knowledge and expertise of research groups from diverse research areas.

The SSH model, an emblem of a one dimensional chiral symmetric TI, that had initially been proposed to explain the striking transport characteristics of {\it trans}-polyacetylenes, is conveniently dealt with in a tight binding formalism describing non-interacting fermions. The arrangement of two bonds, described by two different values of the nearest neighbor hopping (overlap) integrals, gives the lattice a bipartite sublattice structure. The system exhibits nontrivial topological phases enriched with topologically protected edge states, a signature of the TI's. The edge states are robust against defects and disorder, and are fully understood from the single-particle eigenstates.

This remarkable feature inspired a series of experiments in recent past, where several topological networks have been realized and explored, bringing to light a plethora of new physics. The experiments cover a wide canvas of physics, encompassing for example, the exotic bulk properties of the topological character of an SSH chain using real space superlattices,~\cite{folling,sebby} demonstration of robust edge states in  classical systems of coupled mechanical oscillators,~\cite{poli,liu,huber} the dynamics and the topological solitonic states in a momentum-space lattice fabricated with $^{87}$Rb atoms,~\cite{meier} topological Haldane model with  $^{40}$K atoms trapped in a honeycomb lattice,~\cite{jotzu} both in an ultracold-atomic platform, or the experimentally realized symmetry protected topological phase of interacting bosons using Rydberg atoms~\cite{leseleuc} - to name a few. 
Such experiments also received good company in recent times in photonics, where tailor-made {\it lattice geometries} are fabricated on glassy substrates to explore a wide spectrum of issues, ranging from an extreme localization of photonic excitations~\cite{seba1,seba2} to the physics of topological systems, revealing the existence of topological bound states in a Flouquet engineered system resembling the SSH model,~\cite{alex1,rechtsman} or the experimental realization of a topological insulator using photonic Aharonov-Bohm (AB) cages,~\cite{alex2} an important issue in the context of the present article.

\begin{figure}[ht]
\centering
(a) \includegraphics[width=0.75\columnwidth]{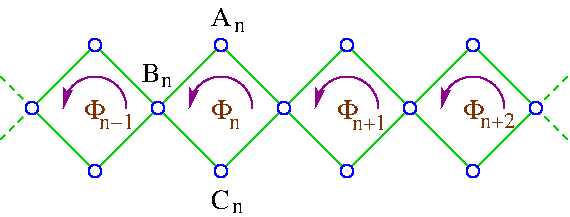} \\
(b) \includegraphics[width=0.75\columnwidth]{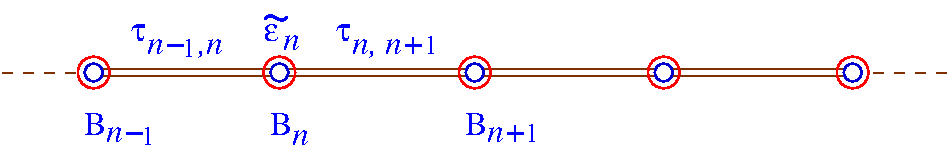} 
\caption{(Color online) 
(a) A quasi-one dimensional endless array of diamond network trapping a non-uniform distribution of magnetic flux, and (b) the \textit{renormalized} chain comprising only the bulk sites $B_n$, and obtained by decimating the vertices $A_n$ and $C_n$ having coordination number equal to two. Here $\tilde{\epsilon}$ and $\tau_{n, n\pm 1}$ are the renormalized on-site potential and the nearest neighbor hopping integrals, in a typical tight binding scheme.}
\label{diamond}
\end{figure}

In the present communication we theoretically investigate the topological properties 
of a flux-staggered array of diamond shaped cells, axially juxtaposed, as illustrated in Fig~\ref{diamond}(a). The out-of plane magnetic field piercing the $n$th cell and the resulting `trapped' flux $\Phi_n$ can have, in principle, any chosen distribution, offering a wide class of staggering patterns. The lattice presented in Fig.~\ref{diamond}(a) is an example of what we shall henceforth refer to as a {\it topological quantum network} (TQN). It is a member of a large group beginning definitely with the paradigmatic SSH chain,~\cite{ssh} and its extensions,~\cite{trimer,miroshnichenko} and include the photonic diamond TQN's,~\cite{alex2} the Lieb lattice models~\cite{jiang,chen,ankita, nathan,ricardo1,ricardo2} or kagome lattices ~\cite{macdonald}  to name a few. In totality, these systems build a formidable (though by no means exhaustive) canvas of studies, shedding light on the intricacies of the band structure and topological order exhibited by the TQN's. 

%%%%%%%%%%%%%%%%%%%%%%%%%%%

Our choice of the diamond-TQN (DTQN) is motivated by certain issues outlined next that we find pertinent.
Firstly, the flux trapped in a cell can be controlled from outside (for example, by controlling the current in a nano-solenoid, that can in principle, be planted in a cell), and this gives us a complete freedom to continuously `deform' the Hamiltonian describing the system through the Peierls' phase associated with the hopping integral, and engineer a {\it topological phase transition} (TPT) almost at will. This, to our mind, is a good choice in many ways as we do not have to tune the values of the intrinsic system-parameters such as the hopping (overlap) integrals like what is done in an SSH case, or control the evanescent coupling of the waveguides in the photonic lattices~\cite{seba1,seba2} or controlling the well-depths in the optical lattice structures,~\cite{fallani} for example. Changing intrinsic parameters essentially means that while experimenting, one has to deal with a different lattice in every occasion. Changing flux from outside gets rid of this ordeal.

Secondly, a linear DTQN threaded by a uniform flux, that has been critically examined before and revealed a wealth of knowledge, such as, a flux-induced semiconducting behavior,~\cite{shreekantha} spin-filtering effects~\cite{aharony1,aharony2} or a spin-selective Aharonov-Casher caging effect for arbitrary spins,~\cite{amrita} has recently turned into an object of supreme interest in state-of-the art photonics, presenting direct experimental evidence of the AB-caging effect.~\cite{seba1,seba2,alex2}
The AB caging, that is an important issue in the present work, was introduced by Vidal et al.~\cite{julien1,julien2} This a geometry-induced {\it extreme localization} of single particle states, seen in periodic lattices. The amplitudes of the wavefunction are localized over a finite number of lattice sites, being zero elsewhere. These `caged' states are entirely degenerate, and are associated with {\it flat} (dispersionless) energy bands (FB) where the energy turns out to be independent of the quasimomentum.~\cite{leykam1}-\cite{rhim} Such FB states, forming a `compacton' structure,  has recently been observed experimentally in an optical waveguide array,~\cite{rodrigo} and are being considered seriously for creation of a whole set of `diffraction-free' modes that can transport quantum information up to a long distance.~\cite{rodrigo,rontgen} The DTQN system thus  serves as a unique platform to study the prospects of a tailor-made quantum matter where one can explore the topological phase transitions, and a geometry-induced extreme localization of single particle states in view of transporting quantum information. 

With this background we present in the following sections the fundamental results obtained and the physics of topological properties exhibited by the DTQN and its unique controllable AB caging aspects. Section II discusses the character of the bulk bands in two elementary variants of the DTQN. In Section III the AB caging and a possible engineering of its spatial extent is talked abut. The physics of the topologically protected edge modes is presented in Section IV, and conclusions are drawn in Section V.
%%%%%%%%%%%%%%%%%%%%%%%%%%%%%%%%%%%%%%%%%%%%%%%%%%%%%%%%%%%%%%%%%%%%%%%%%%%%%%%%%%%%
\begin{figure}[ht]
%\centering
(a)\includegraphics[width=0.60\columnwidth]{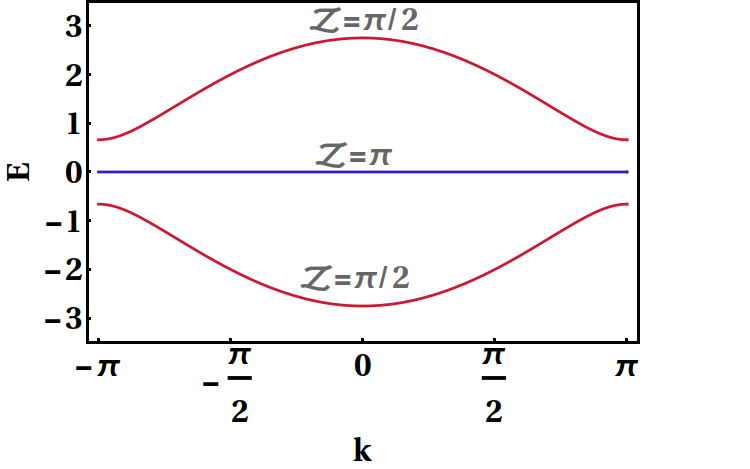}
(b)\includegraphics[width=0.60\columnwidth]{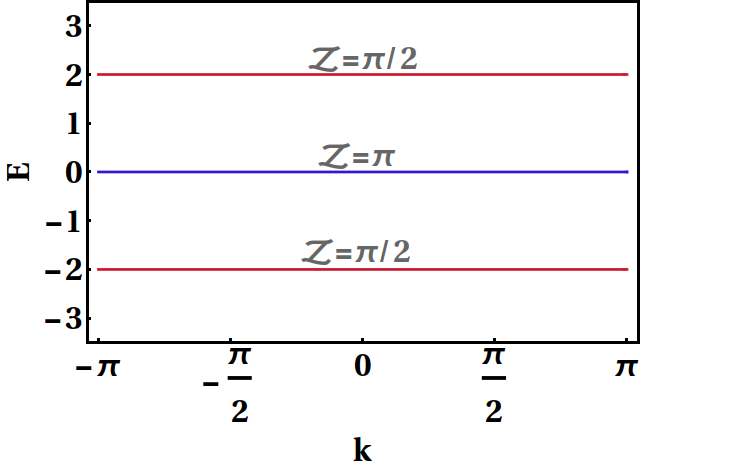}
\caption{(Color online)
Dispersion relation \textit{E(k)} for a diamond network with a constant flux trapped in each cell. (a) $\Phi=0.15 \Phi_0$. Here we get one non-dispersive band at $E=0$, and two dispersive bands. 
(b) $\Phi=0.5 \Phi_0$. Here, the entire spectrum collapses into an extreme localization profile, showing three flat bands at $E = 0$, and $E=\pm 2$. where the energy $E$ is expressed in units of $t$. The on-site potential is set equal to zero for all sites in the DTQN. The Zak-phases in each case are displayed.}
\label{single-band}
\end{figure}

%%%%%%%%%%%%%%%%%%%%%%%%%%%%%%%%%%%%%%%%%%%%%%%%%%%%%%%%%%%%%%%%%%%%%%%%%%%%%
\section{Topology of the bulk bands}
\label{topology}
Our model system is depicted in Fig.~\ref{diamond} (a). 
It is a one dimensional array of diamond shaped cells with infinite axial span. The $n$-th cell traps a finite magnetic flux $\Phi_{n}$.  
We describe the system by the standard tight-binding Hamiltonian 
for non-interacting spinless electrons, written in the Wannier basis, viz., 
\begin{equation}
H =  \sum_{<jl>} (|c_j \rangle t_{jl} e^{i \theta_n} \langle c_l| + h.c.) + \sum_{j} |c_j \rangle \epsilon_{j} \langle c_j|
\label{ham}
\end{equation}
The first term is the nearest neighbor hopping integral parameter that physically bears the kinetic signature and the rest part denotes the on-site potential energy of the electron at the respective atomic sites. In this communication, we will set $\epsilon_j=\epsilon=0$ and $t_{jl}=t=1$ throughout. The exponential factor tagged to the hopping integral is the Peierls' phase factor. Here, $\theta_n = 2 \pi \Phi_{n} / 4 \Phi_{0}$ ($\Phi_{0} = h c/e$ being the elementary flux quantum). We work in a gauge where $\theta_n$ is same for every pair of nearest neighboring sites in any $n$-th cell, and is directly proportional {\it only} to the flux $\Phi_n$ trapped in the $n$-th cell.
Though the atomic sites have coordination numbers two (marked as $A_n$ and $C_n$) or four (marked as $B_n$), we set, without any loss of generality, the on-site potential for every site equal to zero throughout the analysis. 

Here we draw the reader's attention to a pertinent and an interesting issue. The magnetic flux changes from one diamond unit to the next. As flux lines are related to the external magnetic field by $\Phi_n = \mathbf{B} \cdot \mathbf{A} = B A \cos \gamma_n$, $\gamma_n$ being the 
angle between the field and the area vector $\mathbf{A}$ of the plaquette, 
the physics offered by the general staggered flux pattern should be equivalent to that of the same diamond array, placed in a uniform magnetic field, but now with its plaquettes rotated around the major axis at arbitrary angles. For a discrete distribution of flux, the angles of twist will naturally be discrete, but the present model allows us to explore an even more general situation of a continuous axial twist, as we will discuss in Section III .

%%%%%%%%%%%%%%%%%%%%%%%%%%%%%%%%%%%%%%%%%%%%%%%%%%%%%%%%%%%%%%%%%%%%%%%
\subsection{Uniform flux distribution: $\Phi_n=\Phi$} 
\label{uniform}
%%%%%%%%%%%%%%%%%%%%%%%%%%%%%%%%%%%%%%%%%%%%%%%%%%%%%%%%%%%%
\begin{figure}[ht]
(a)\includegraphics[width=0.5\columnwidth]{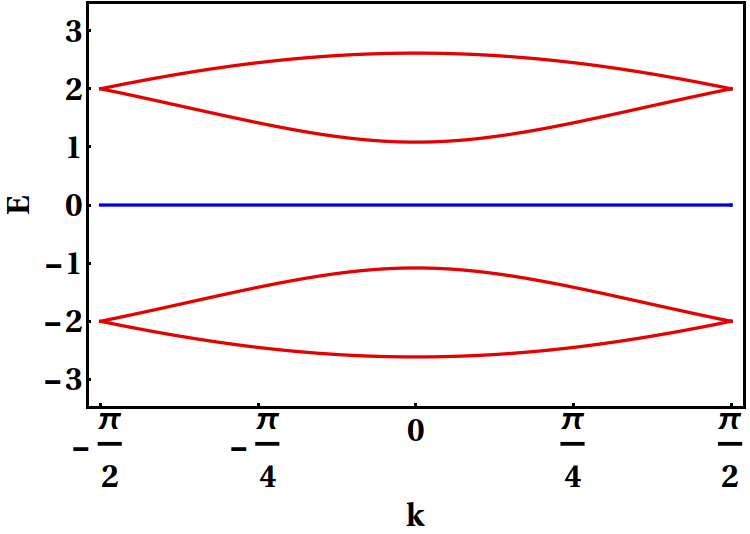}\\
(b)\includegraphics[width=0.5\columnwidth]{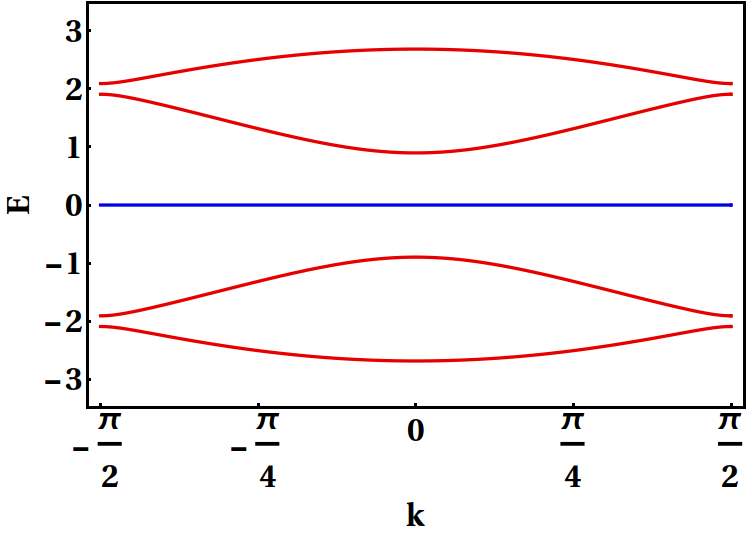}\\
(c)\includegraphics[width=0.5\columnwidth]{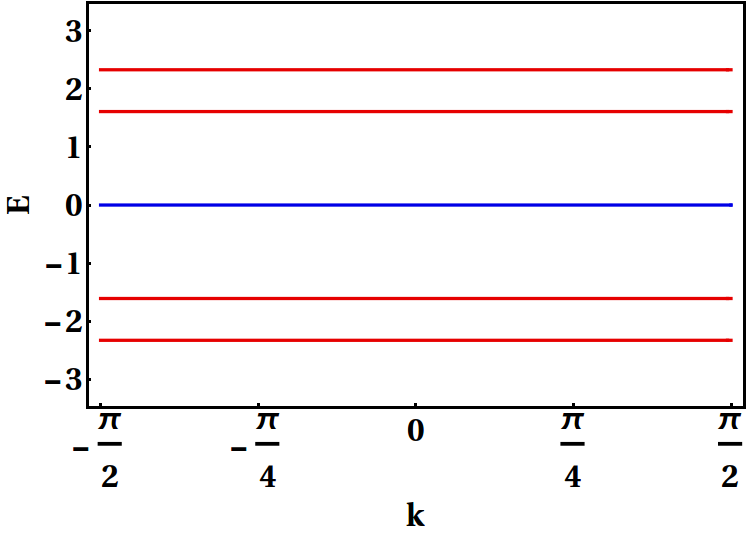}
\caption{(Color online) 
Band dispersion (energy-momentum relationship) for a diamond array with periodic binary flux distribution. (a) $ \Phi_1 = \Phi_2 = 0.25 \Phi_0$, (b) $ \Phi_1 = 0.25 \Phi_0$, $\Phi_2 = 0.15 \Phi_0$, and (c) $\Phi_1 = 0.25 \Phi_0$, $\Phi_2 = 0.5\Phi_0 $.  Lattice periodicity $a$ is set as unity in the numerical calculation.}

\label{binary}
\end{figure}
%%%%%%%%%%%%%%%%%%%%%%%%%%%%%%%%%%%%%%%%%%%%%%%%%%%%%%%%%%%%
One can recast the Hamiltonian in the momentum space using the following relation
\begin{equation}
H = \sum_{k} \psi^{\dagger}_{k} \mathcal{H} (\bm{k}) \psi_{k}
\label{ktransform}
\end{equation}
where, 
%\begin{widetext}
\begin{eqnarray}
\mathcal{H} (k) =  
\left( \begin{array}{cccc}
0 & t_{B} + t_{F} e^{-i k a} & t_F + t_B e^{-i k a}\\ 
t_F + t_B e^{i k a} & 0 & 0\\ 
t_B + t_F e^{i k a} & 0 & 0
\end{array}
\right) ~
\label{kspace}
\end{eqnarray}
%\end{widetext}
where, $t_{F(B)}=t e^{\pm i \theta}$ and $\theta=2\pi\Phi/4\Phi_0$, $k$ is the wave vector, and $a$ is the lattice periodicity. 
%-------------------------------------------------------%
On diagonalizing $\mathcal{H}(k)$ the band dispersion in terms of applied flux $\Phi$ turns out to be, $E =0$, $\pm 2 t \sqrt{1+\cos (\pi \Phi/\Phi_0) \cos ka}$. The `band' at $E=0$ is a flux-insensitive robust and non-dispersive one, representing a {\it flat band} (FB) state~\cite{leykam1,travkin,leykam2} at the band center, and the other two represent the two flux-dependent dispersive bands. The spectrum turns out to be gap-less for $\Phi=0$ and $\Phi=\Phi_0$, while a gap at the Brillouin zone boundary opens and remains open, for any non-zero value of the trapped flux. 

The topological character of the bands is customarily studied through a topological invariant, which is the Zak-phase~\cite{zak} in this case.
This quantity, essentially similar Berry's phase,~\cite{berry} can be written in terms of the $j$-th Bloch eigenstate as,
\begin{equation}
\mathcal{Z} = \frac{i}{2 \pi} \int_{0} ^{2 \pi} d k \langle u_{k,j}| \partial_{k} u_{k,j} \rangle
\label{zakeq}
\end{equation}
Experimental measurement of Zak phase in photonic counterparts of similar topological network has already been accomplished~\cite{atala}.

We have evaluated the Zak-phase using the Wilson-loop approach~\cite{fukui} that removes the gauge dependence of $Z$. The integration of Eq.~\eqref{zakeq} is converted into a summation over the entire Brillouin Zone, divided into $N$ number of identical discrete segments such that each interval in wave vector becomes $\Delta k = 2 \pi / N a$, $a$ being the lattice constant. We take $N=400$ which ensures a convergence, and a reasonably good approximation to the original integral. In this discrete notation, within the  Wilson loop prescription the Zak-phase is expressed, for a non-degenerate $s$-th band, as,~\cite{fukui,marzari}
\begin{equation}
\mathcal{Z} = -Im \left( \log \prod_{k_i} \langle u_{k_{i},s}| u_{k_{i+1},s} \rangle \right)
\label{zakf2}
\end{equation}
As illustrated in Fig.~\ref{single-band} both the dispersive bands display non-quantized Zak-phase (attributed to the broken translational invariance due to the magnetic field) equal to $\pi/2$ and the central flat band a quantized value $\pi$.~\cite{alex2} The values of $Z$ are found to be independent of the magnetic flux, even in the limit $\Phi=\Phi_0/2$, when the entire spectrum collapses into just three non-dispersive bands and the system displays an extreme localization. This is the most elementary distribution of an AB-caging, as will be discussed in the next section.

\subsection{The simplest flux-staggered array}
The simplest staggered flux distribution is achieved by periodically repeating two different magnetic flux values, viz., $\Phi_1$ and $\Phi_2$ in the consecutive cells of an infinite array. The Hamiltonian matrix in this case has a dimension $6 \times 6$, and is given by, 
\begin{widetext}
\begin{equation}
\mathcal{H}(\bm{k})=
\left[ \begin{array}{cccccc}
0 & t_{B}(1) & t_{F}(1) & 0 & t_{F}(2)e^{-2ika} & t_{B}(2)e^{-2ika} \\
t_{F}(1) & 0 & 0 & t_{B}(1) & 0 & 0 \\
t_{B}(1) & 0 & 0 & t_{F}(1) & 0 & 0 \\
0 & t_{F}(1) & t_{B}(1) & 0 & t_{F}(2) & t_{B}(2) \\
t_{B}(2)e^{2ika} & 0 & 0 & t_{F}(2) & 0 & 0\\
t_{F}(2)e^{2ika} & 0 & 0 & t_{B}(2) & 0 & 0
\end{array}
\right]
\end{equation}
\end{widetext}
where, $t_{F(B)}(1)=t \exp ({\pm i \pi \Phi_1/2 \Phi_0})$ and  
$t_{F(B)}(2)=t \exp ({\pm i \pi \Phi_2/2 \Phi_0})$.
Straightforward diagonalization of the above matrix provides the entire band diagram. The dispersion relations are $E=0$ (a two-fold degenerate band) and 
\begin{widetext}
\begin{equation}
E=\pm \sqrt{(4 \pm \sqrt{2} \sqrt{(2+ \cos (2 \pi \Phi_1 / \Phi_0)+\cos (2 \pi \Phi_2 / \Phi_0)+ 
F_{1} + 
F_{2} + 
F_{3} +
F_{4}})} 
\end{equation}
\end{widetext}
where, $F_{1} = \cos [2 ka - (\pi/\Phi_0) (\Phi_1+\Phi_2)]$,
$F_{2} = \cos [2 ka + (\pi/\Phi_0) (\Phi_1+\Phi_2)]$,
$F_{3} = \cos [2 ka - (\pi/\Phi_0) (\Phi_1-\Phi_2)]$ and
$F_{4} = \cos [2 ka + (\pi/\Phi_0) (\Phi_1-\Phi_2)]$. 
The band at $E=0$ does not show any response to the applied perturbation but the four other bands have the obvious flux sensitivity. Fig.~\ref{binary} displays a subset of the dispersion relations for certain special values of the magnetic flux, and is going to play an important role in the subsequent discussion.
%%%%%%%%%%%%%%%%%%%%%%%%% Wilson Loop %%%%%%%%%%%%%%%%%%%%%
We have calculated the Zak phase in this case of staggered flux geometry. There is degeneracy in the central band. Following the Wilson loop approach for degenerate bands we replace Eq.~\eqref{zakf2} by a product of `overlap matrices'~\cite{overarxiv}, viz, 
\begin{equation}
\mathcal{Z}(k_i)=-Im \left[ \log \left( \prod_{k_{j}} S_{(k_{i},k_{j}),(k_{i},k_{j+1})} \right) \right]
\end{equation}
where, $\boldmath{S}$ is the overlap matrix~\cite{overarxiv}.
%%%%%%%%%%%%%%%%%%%%%%%%%%%%%%%%%%%%%%%%%%%%%%%%%%%%%%%%%%%

For a binary flux staggered lattice no quantized Zak phase is found. The values are scattered, without showing any definite correlation. However, the same values are obtained for symmetrically distributed energy eigenvalues $\pm E$. For example, with $\Phi_1=0.25 \Phi_0$ and $\Phi_2=0.15 \Phi_0$ the energy eigenvalues are $E=0$, $\pm 1.90561$ and $\pm 2.09013$ in units of the hopping integral $t$ (set equal to unity). The corresponding Zak phase values are obtained as, $Z=0.888 \pi$, $0.851 \pi$ and $-0.241 \pi$ respectively. The values keep changing as the flux patterns are made to vary, retrieving the old value of $\mathcal{Z} = \pi$ at $E=0$. At this point it becomes a pertinent question to ask ourselves whether there can still be a topological order, or a TPT even if the system doesn't show up any time reversal symmetry or any definitive change in the values of the Zak phase.

We find that, it does really. To address this issue, in this subsection we take a different approach, based on a real space decimation of a subset of the vertices (these are $A$ and $C$-type sites) in the original diamond array to map it onto an effective SSH chain, where we argue that a {\it topological phase transition} (TPT) can be initiated by tuning the values of the magnetic flux. As we mentioned in the introduction, the flux can be tuned from {\it outside} and is very much in the hands of an experimentalist. 

Consider the $n$-th cell. The difference equation connecting any $j$th site with its nearest neighbors $l$ in cell number $n$, is given by,
\begin{equation}
(E - \epsilon) ~\psi_j = t\sum_{l}  e^{\pm i \theta_n} ~\psi_{l}
\label{diff1d}
\end{equation}
where, $\theta_n =2 \pi \Phi_{n}/4\Phi_0$ ($n=1$, $2$).
Let us specifically focus at the physics when the energy $E \ne 0$. This will suffice.
Using Eq.~\eqref{diff1d} we decimate the top
and the down ($A_n$ and $C_n$) vertices (that is, we eliminate the amplitudes $\psi_{A_n}$ and $\psi_{C_n}$ and retain only the amplitudes $\psi_{B_n}$) to map the original DTQN array into an effective one-dimensional chain, described by a difference equation,
\begin{equation} 
[(E-\epsilon)^2 - 4t^2]~ \psi_{B_n} = 2t^2 \cos 2\theta_1 ~\psi_{B_{n-1}} + 2t^2 \cos 2\theta_2 ~\psi_{B_{n+1}}
\label{rgeqn}
\end{equation} 
One can easily check~\cite{shreekantha} that this equation yields the exact density of states (DOS) profile of the original diamond chain barring the isolated spike at $E=0$, the latter being a contribution from the top vertices $A_n$ and $C_n$. But as we shall see, it is enough to get a flavour of a TPT from this equation. Eq.~\eqref{rgeqn} can be written as, 
\begin{equation}
E'\psi_{B_n} = \tau_1\psi_{B_{n-1}} + \tau_2 \psi_{B_{n+1}}
\label{ssh}
\end{equation}
with $E'=[(E-\epsilon)^2 - 4t^2]$, and $\tau_n=2t^2 \cos 2\theta_n$ ($n=1$, $2$). Eq.~\eqref{ssh} describes an SSH chain where the `on-site' potential is {\it zero}, and the hopping integrals $\tau_1$ and $\tau_2$ alternate periodically. 

The decimation reduces the number of degrees of freedom to deal with. The effective SSH chain, now characterized by two hopping integrals $\tau_1=2t^2 \cos 2\theta_1$, and $\tau_2=2t^2 \cos 2\theta_2$ is thus defined on a {\it reduced} (real) subspace (spanned by a lesser number of the degrees of freedom). From this angle, it can be taken to describe  a `square root' topological system.~\cite{alex2,arkinstall} This direct equivalence with the basic SSH chain implies that, if we set $\epsilon=0$ and $t=1$ in the parent diamond chain, the opening and closing of energy gaps will be observed at $E'=0$, which amounts to $E=\pm 2$ in terms of the eigenvalues of the original DTQN. Fig.~\ref{binary} (a) and (b) display the closed gap (when $\Phi_1=\Phi_2$), and its opening (for $\Phi_1 \ne \Phi_2$) at the Brillouin zone boundaries (now at $\pm \pi/2$ owing to the double period).

Eq.~\eqref{ssh} can be thought to be obtained from a Hamiltonian $\mathbf{H}^\ast$ of a bipartite lattice of length $L$ say, defined on this `reduced degrees of freedom' space as,

\begin{eqnarray}
\mathbf{H}^\ast =  
\left( \begin{array}{cc}
0 & H_{ab} \\ 
H^\dag_{ab} & 0
\end{array}
\right)
\label{hamssh}
\end{eqnarray}
Here, $a$ and $b$ refer to the two kinds of vertices, sitting in between the pairs of hopping $\tau_1-\tau_2$ and $\tau_2-\tau_1$ respectively, in the effective one dimensional bipartite lattice (Fig.~\ref{diamond}(b), with $\tau_{n-1,n}=\tau_1$ and $\tau_{n,n+1}=\tau_2$, and repeating periodically). The mapping obviously relates the topological properties of the Hamiltonian $\mathbf{H^\ast}$, that now exhibits a chiral symmetry $\sigma_z \mathbf{H^\ast}\sigma_z = - \mathbf{H^\ast}$, with that shown by the standard SSH model. Since the eigenstates of $\mathbf{H}^\ast$ are also the eigenstates of the original Hamiltonian $\mathbf{H}$, the states with $E = \pm 2$ are protected by chirality. It will be argued in the next section that a cut-piece of a diamond network with a binary flux distribution can indeed have edge-localized states around $E=\pm 2$ (along with other energies, depending on the relative flux ratio). Such edge localized states will thus also be protected by chiral symmetry.

The {\it effective} SSH chain has hopping integrals that are engineered by the applied magnetic field only. It can be easily checked (and has been depicted in Fig.~\ref{binary}(a) and (b)) that, the gap around $E=\pm 2$ can be continuously closed and re-opened on simultaneous tuning of $\Phi_1$ and $\Phi_2$. This indicates a TPT. We refer the reader to the Appendix for details.

Before ending this section, it may be noted that higher order staggering effects can easily be dealt with now. Of course, a study of the bulk band properties needs handling of larger unit cell matrices. We skip such details here to save space, though a tertiary flux distribution will be discussed in the context of AB caging and the edge states in the subsequent sections.

\section{Selective Aharonov-Bohm caging}
\label{abcaging}
\subsection{The easiest caging with uniform flux}
The diamond array considered here exhibits
%===========================================================================
\begin{figure}[ht]
\centering
(a) \includegraphics[width=0.35\columnwidth]{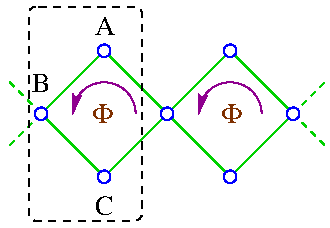}
(b)\includegraphics[width=0.35\columnwidth]{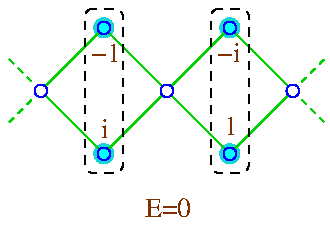}
(c)\includegraphics[width=0.35\columnwidth]{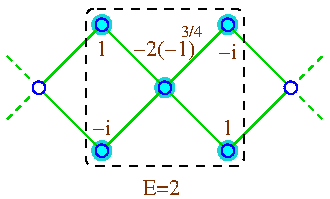}
(d)\includegraphics[width=0.35\columnwidth]{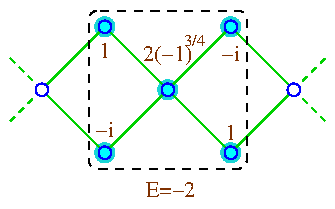}
\caption{(Color online) 
(a) Schematic demonstration of an infinite array of diamond network threaded with uniform magnetic flux. The unit cell consists of three atomic sites labelled as A, B and C. 
(b) $U (1)$ class of compact localized state at $E=0$. This is flux-independent.
(c) and (d) represent a $U (2)$ class of compact localized state. This is a quincunx profile at the special value of $\Phi=0.5\Phi_0$, and at $E=\pm2$ respectively. The unit cell and the CLS' s are marked by the rectangular dashed line with the locations of the non-zero amplitudes marked deep blue.}
\label{caging-single-flux}
\end{figure}

%%%%%%%%%%%%%%%%%%%%%%%%%%%%%%%%%%%%%%%%%%%%%%%%%%%%%%%%%%%%%%%%%%%%%%%%%
AB-caging.~\cite{julien1,julien2} This caging effect, in its simplest form, manifests itself through local, clustered distribution of amplitudes of the wave function. These are the so called {\it compact localized states} (CLS).~\cite{maimaiti,flach1,flach2} The consequential appearance of flat, non-dispersive bands (three in number when the flux is uniform throughout) marks the phenomenon. For the simplest caging, the bands are at $E=0$, and $E=\pm 2$.

The non-dispersive band at $E=0$ is independent of the choice of the magnetic flux. The 
only possibility to satisfy the Schr\"{o}dinger equation is to make the amplitude of the wave function vanish at the bulk $B_n$ sites. One such configuration is shown in Fig.~\ref{caging-single-flux}(b). This is the simplest example of a CLS and AB-caging, the strongest of the $U(1)$ class of CLS.~\cite{leykam1} The amplitudes are confined within one unit cell of the DTQN. The states at $E=0$ are robust, and topologically protected, characterized by a quantized Zak phase $\pi$.

In the case where $E \ne 0$, we observe that, as the flux per plaquette is set at a uniform value $\Phi=\Phi_0/2$, the effective hopping integral, connecting a bulk ($B_n$) site to its nearest neighbors becomes {\it zero}. This can be directly observed by decimating the top ($A_n$) and bottom ($C_n$) vertices using Eq.~\eqref{diff1d}. The renormalized hopping on the effective one dimensional chain becomes equal to $t_{eff}=[2t^2 \cos (\pi\Phi/\Phi_0)]/(E-\epsilon)$, and becomes {\it zero} at $\Phi=\Phi_0/2$. This is the basic reason underlying an extreme localization in such cases, and the formation of the CLS's. As for the flat-band energy eigenvalues $E=\pm 2$,
and $\Phi=\Phi_0/2$, Eq.~\eqref{diff1d} is easily solved to find a quincunx distribution of amplitudes, spread over a set of five vertices spanning two consecutive unit cells. This is a larger patch of non-zero amplitudes, and the CLS is in a class $U(2)$.~\cite{leykam1} Needless to say, in both the cases, one island of non-zero amplitudes is separated from the neighboring islands by sites, where the wave function has a vanishing amplitude. The $U(2)$ class of CLS is depicted in Fig.~\ref{caging-single-flux} (c) and (d) for $E=\pm2$.

%%%%%%%%%%%%%%%%%%%%%%%%%%%%%%%%%%%%%%%%%%%%%%%%%%%%%%%%%%%%%%%%%%%%%%%%%%%%%%%%%
\begin{figure}[ht]
\centering
(a)\includegraphics[width=0.85\columnwidth]{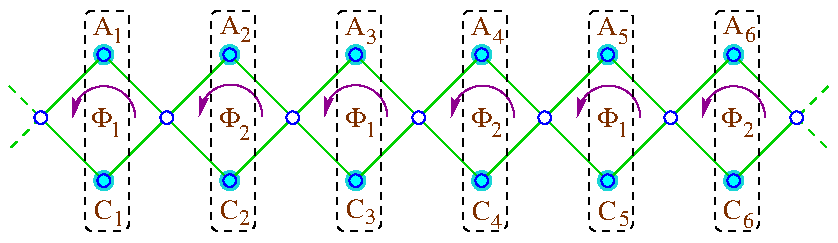}\\
(b)\includegraphics[width=0.85\columnwidth]{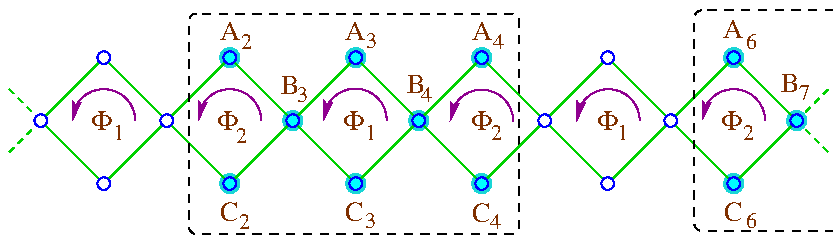}\\
\caption{(Color online) Aharonov-Bohm caging in binary (a), (b), flux staggered lattices. The non-vanishing amplitudes are imprinted by bright cyan dots. The caging, self explanatory, is bordered by dashed boxes. In (b) the distribution shown is typically obtained by setting $\Phi_2=\Phi_0/2$, as described in the text. }
\label{bispec}
\end{figure}
%============================================================================
\subsection{Extending the quantum prison}
The AB-caging at the centre of the band, i.e. for $E=0$, is independent of the flux, in general. But, the vanishing of the effective hopping connecting the $B_n$ vertices on a renormalized linear chain (Fig.~\ref{diamond}(b)), whenever $\Phi_n=\Phi_0/2$, opens up a possibility of engineering the `physical spread' (in real space) of the amplitudes of a CLS when the staggered flux distribution has a higher order periodicity. One such extended distribution, for a binary ordered staggering with $\Phi_1$ and $\Phi_2$ repeated in an alternate fashion, is explicitly worked out and illustrated in Fig.~\ref{bispec}, where the sites with non-zero amplitudes are colored brightly.
 
 In the binary flux staggered case the easiest (and strongest) AB-caging is again observed for $E=0$. The distribution of amplitudes of the wave function can be worked out analytically. The amplitudes of the wave function are pinned at the top ($A_n$) and bottom ($C_n$) sites of any $n$th diamond cell. The amplitude at every $B_n$ site having a coordination number of four is zero. A special situation can be cited as an example. We refer to $E=0$. We can construct $\psi_{j}=-(\cos\theta_2/\cos \theta_1)$ for $j\subset (A_1,C_1)$, and $\psi_j=1$ for $j \subset (A_2, C_2)$ in Fig.~\ref{bispec}(a), and repeating periodically. This is a $U(1)$ class of CLS again. However, in this special example with $E=0$, one needs to assure that, $\Phi_1 \ne \Phi_0$. Other degenerate solutions can of course, be worked out. We do not provide all of these, just to save space.
 
 The $U(2)$ class of CLS for a $\Phi_1$-$\Phi_2$ periodic distribution can be worked out by hand as well. A specific distribution is depicted in Fig~\ref{bispec}(b) for $\Phi_2=\Phi_0/2$. 
 For such a flux pattern, we can always work out a particular solution, satisfying Eq.~\ref{diff1d}. For example, let $\psi_{A_2}=\psi_{C_4}= x e^{-i\theta_2}$, $\psi_{C_2}=\psi_{A_4}=x e^{i\theta_2}$, $\psi_{A_3}=\psi_{C_3}=y$, and $\psi_{B_3}=\psi_{B_4}=z$. If $z=1$ is chosen arbitrarily, so that both $x$ and $y$ are evaluated in unit of $z$, then one can easily work out a solution for 
 $E=\pm \sqrt{2+4 \cos^2 \theta_1}$. Both roots are real, and a $\Phi_1$- dependent set of amplitudes $x=1/E$, and $y=(2/E) \cos \theta_1$, in complete agreement with Eq.~\eqref{diff1d}, can be obtained. The dependence of the energy eigenvalue and thus, the amplitudes on $\theta_1$, i.e. on $\Phi_1$ exposes the options of tuning the extreme localized states at different energies, keeping the pattern of the distribution intact.
 
 With a tertiary flux ordering, one has more options of tuning the relative magnitudes of the trapped fluxes, and generate both $U(1)$ and $U(2)$ classes of CLS. It can be easily understood now that, the $U(1)$ class will have a distribution where the amplitudes are pinned at the $(A_n, C_n)$ vertices for $E=0$. The $U(2)$ class of CLS spreads over a unit cell, that now comprises three consecutive cells trapping fluxes $\Phi_1$, $\Phi_2$ and $\Phi_3$. We have obtained a variety of distributions, but refrain from overloading the text with these, as the basic idea, to our mind is conveyed already. As expected, with an increased staggering period, the spatial extent of a unit cell increases and the caging spreads over larger patches of the DTQN.
 
%%%%%%%%%%%%%%%%%%%%%%%%%%%%%%%%%%%%
% UNIFORM FLUX EDGE PIC
\begin{figure*}[ht]
(a)\includegraphics[width=0.75\columnwidth]{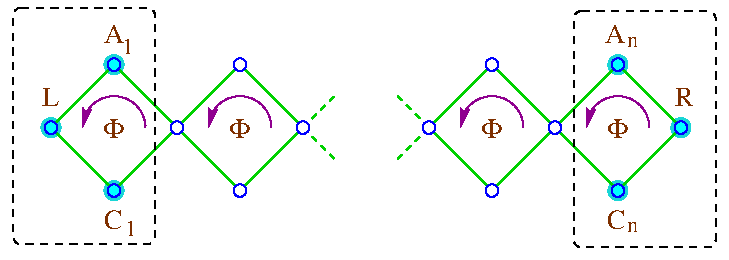}
%(b)\includegraphics[width=0.45\columnwidth]{edge-ampli-single-2.png}\\
(b)\includegraphics[width=0.65\columnwidth]{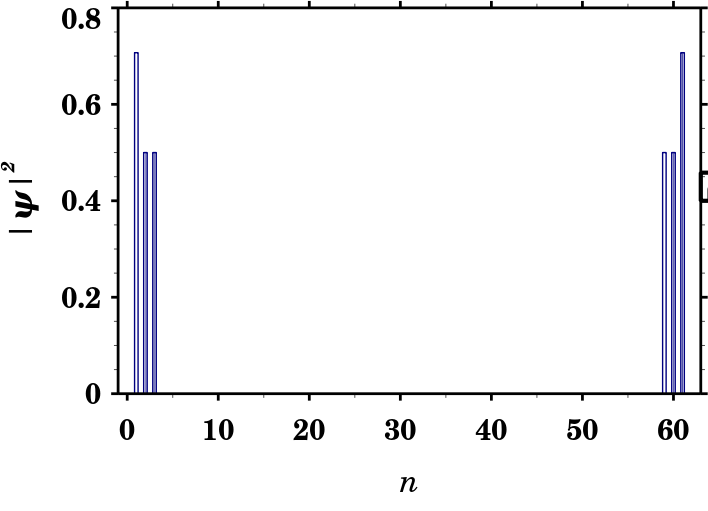}\\
(c)\includegraphics[width=0.70\columnwidth]{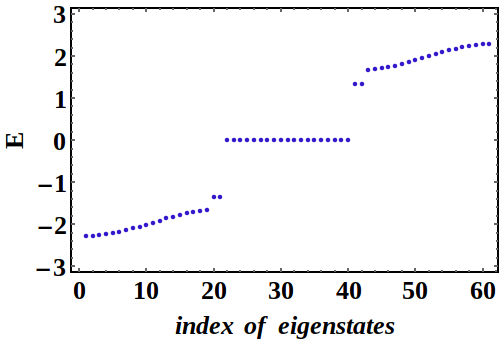}
%(b)\includegraphics[width=0.45\columnwidth]{edge-ampli-single-2.png}\\
(d)\includegraphics[width=0.65\columnwidth]{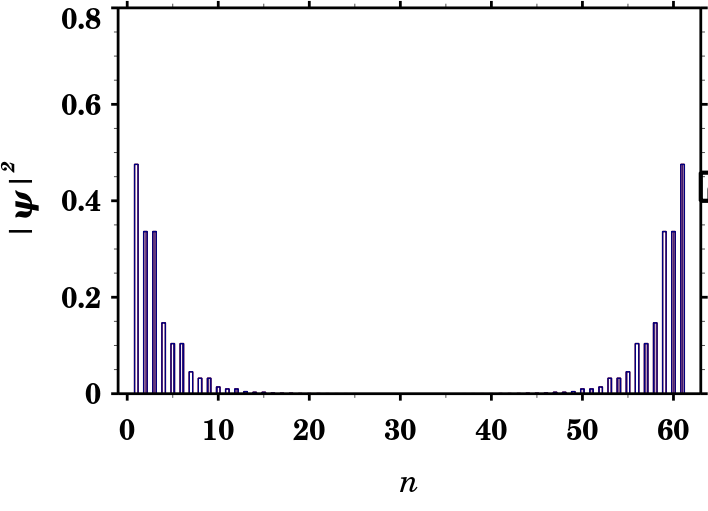}
\caption{(Color online) 
Edge state-amplitudes for a finite diamond array threaded by a constant flux $\Phi =\Phi_0/2$. (a) At $E=\sqrt{2}$ the amplitudes are pinned over a triplet of sites, with $\psi_{A_1} = \psi_{C_n}=(1+i)/2$, $\psi_L =\psi_R=1$ and $\psi_{C_1}=\psi_{A_n} = (1-i)/2$. A similar distribution for the edge state can be obtained for $E=-\sqrt{2}$. (b) Plot of probability density for the state at (a).
(c) A `non-half flux' case, $\Phi=0.6\Phi_0$. Here, $E=\pm 1.345$. 
The edge states are again at the band gaps, and are supported by special pairs of eigenstates as shown. (d) The amplitudes corresponding to the edge states in (c), are localized and clustered over a small finite set of sites at both the edges.}
\label{edge}
\end{figure*}
%%%%%%%%%%%%%%%%%%%%%%%%%%%%%%%%%%%%
\subsection{Engineered caging in a modulated flux profile}

The fact that, setting $\Phi_n=\Phi_0/2$ in any $n$-th cell can make the effective, renormalized coupling between the bulk sites $B_n$ vanish,  one can, in principle, design the caging at will. To implement this, we propose an Aubry-Andr\'{e}-Harper (AAH) modulation~\cite{aubry,harper,thouless} in the flux distribution, given by,  $\Phi_n/\Phi_0 = \lambda \cos (\pi Q n^\alpha a)$. $\lambda$ is the strength of the modulation and the exponent $\alpha$ controls the {\it slowness} of variation of the profile. In the special case of $Q$ being an irrational number, the period of modulation becomes incommensurate with the underlying lattice-period. This particular form of modulations has so far been widely explored in relation to the localization issues in deterministically disordered systems,~\cite{sankar1,sankar2,sankar3} and recently in the study of topological modes in a one dimensional modulated potential profile.~\cite{sankar4}.

In the present proposition, the AAH modulation plays a different role. It determines the staggering profile of the flux. As can easily be understood, the above distribution of magnetic flux is completely equivalent to an infinite diamond array, immersed in a uniform magnetic field, and having an {\it axial twist} through an $n$-dependent angle $\gamma_n = \pi Q n^\alpha a$. The effective hopping between the neighboring bulk sites $B_n$ is now equal to $[2 t^2 /(E-\epsilon)]\cos[\pi \lambda \cos (\pi Q n^\alpha a)]$, and becomes equal to zero for 
\begin{equation}
    \cos (\pi Q n^\alpha a) = \frac{1}{2\lambda}
\label{antiresonance}
\end{equation}

%%%%%%%%%%%%%%%%%%%%%%%%%%%%%%%%%%%%%%%%%%%%%%%%%%%%%%%%%%%%%%%%%%%%%%%%%%%%%%%%%%%
This in fact, sets a threshold for the flux window $\lambda$, viz, $|2\lambda| \ge 1$. For a given selection of $\lambda$, the choices of a rational value of $Q$ and the `slowness' parameter $\alpha$, which amounts to a definite set of the angles of axial twist, determine the plaquette number $n$ in which the `effective', `end-to-end' coupling between the bulk ($B_n$) sites disappears, leading to an AB caging. 
It is not unnatural, in fact it is definitely possible that, the amplitudes remain non-zero on all (or at least a majority of) the cells preceding the $n$-th cell, and after it. This pattern will be repeated as the geometry is periodic as long as $Q$ is a rational number. Thus, the effective area of the quantum prison is extended beyond just a single plaquette, and that too in an engineered way. Extensive numerical search has revealed that cutting off the effective connection in the $n$th plaquette leads to the formation of $2n+1$ flat bands. These non-dispersive bands crowd towards the edges of the energy spectrum, as observed. 

Emphasizing on the equivalence of the AAH modulation with a continuous axial twist, we bring to the notice of the reader this extremely interesting and 
non-trivial {\it tilt induced selectivity} in the Aharonov-Bohm caging and a lazy extreme localization (dictated by the choice of rational $Q$-values). The problem seems challenging from an experimental perspective.
%%%%%%%%%%%%%%%%%%%%%%%%%%%%%%%%%%%%%%%%%%%%%%

\section{TOPOLOGICAL EDGE MODES}
\subsection{A finite array with uniform flux}
In Fig.~\ref{edge} (a) we exhibit a finite diamond array with each cell pierced by the same magnetic flux $\Phi$. The extreme sites to the left and to the right of the array are marked $L$ and $R$ respectively. Here we observe that when we set $\Phi=\Phi_0/2$,  a pair of topologically non-trivial edge modes at $E=\pm \sqrt{2}$ appear in the  spectrum.~\cite{alex2} Each member of the pair shows amplitudes distributed over just three vertices (dotted cage with solid blue sites) around each edge.  The distribution is robust against a physical distortion, as has been  verified by introducing disorder in the distribution of the hopping integrals. The probability distribution of this sharp pinning of states is shown in Fig.~\ref{edge}(b), and is at per with (a).

The paired edge states can also appear for non-half flux cases, as shown in Fig.~\ref{edge}(c), where $\Phi=(3/5) \Phi_0$, and $E=\pm 1.345$. The results are obtained on diagonalizing a twenty ($20$)-cell long finite DTQN. Degenerate pairs of gap states, the $20$th, $21$st and the $40$th and $41$st ones in an array of sixty one ($61$) single particle states supported by the entire array are clearly seen in Fig.~\ref{edge}(c), and their localized profile is depicted in  Fig.~\ref{edge}(d).

%%%%%%%%%%%%%%%%%%%%%%%%%%%%%%%%%%%%%%%%%%%%%%%
\begin{figure}[ht]
\includegraphics[width=0.95\columnwidth]{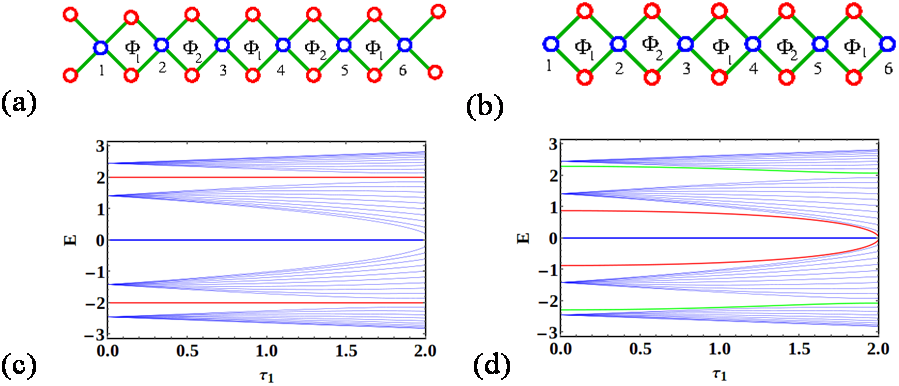}
\caption{(Color online) 
Spectral profile as a function of $\tau_1$(defined as $2\cos (\pi\Phi_1/\Phi_0)$ ) for a finite diamond chain with binary flux distribution at $\Phi_2 = 0$ that makes $\tau_2=2$. 
(a) A pair of dangling atoms are attached at each of the two extreme ends of a finite diamond chain with $20$ cells. Here, the two in-gap edge states are shown in red color at $E=\pm2$, and shown in (c).
(b) The $20$ cell long diamond array without any dangling bonds. There are now
four in-gap edge states that are marked by green and red color respectively, as depicted in panel (d).}
\label{binary-ssh}
\end{figure}
%%%%%%%%%%%%%%%%%%%%%%%%%%%%%%%%%%%%%%%%%%%%%%%%
\subsection{The binary flux-staggered array}
In a conventional, finite SSH chain, comprising two kinds of hopping integrals $\tau_1$ and $\tau_2$ alternating periodically, and beginning with $\tau_1$ (without losing any generality), a topologically non-trivial extreme {\it dimerized} situation is created~\cite{asboth} by setting $\tau_1=0$, and $\tau_2 \ne 0$. The on-site potential $\epsilon$ for a purely one dimensional SSH model is conveniently set equal to zero for {\it all sites}, leading to the immediate conclusion that a topologically protected  edge state will appear at an energy $E=0$. In our case, we argue that, 
the localized edge modes in a finite binary-flux staggered DTQN can easily be understood by mapping it onto an effective SSH chain. This needs a one step renormalization process, through the decimation of the top ($A_n$) and the bottom ($C_n$) vertices. This is easily accomplished by Eq.~\eqref{rgeqn}. We discuss two cases to put forward our idea.

\subsubsection{A finite DTQN with a pair of dangling atoms at each end}
Let us first consider Fig.~\ref{binary-ssh}(a) first, where a finite DTQN has a pair of dangling atoms (or equivalently, bonds) at each end. We have two distinct flux values $\Phi_1$ and $\Phi_2$ trapped periodically in this network. Without losing any generality, we assume the first plaquette to have flux $\Phi_1$. As described in Section~\ref{topology}, after a square-root mapping, the renormalized DTQN chain  resembles an SSH lattice, described by Eq.~\eqref{rgeqn}, or equivalently, by Eq.~\eqref{ssh}. One identifies $E'=(E-\epsilon)^2 - 4t^2$ and the SSH hoppings  $\tau_n = 2\cos 2\theta_n$ for $n=1$,$2$, and with $\theta_n=\pi\Phi_n/2\Phi_0$, as already explained in Eq.~\eqref{ssh}. This mapping immediately tempts us to believe that, in this finite DTQN where all the sites are  equivalent (each `originating' from the parent $B_n$ sites having a coordination number four), we are going to find localized edge-modes at $E'=0$. The answer is in the affirmative, but now it is sensitive to and constrained by the relative strengths of the flux trapped in the consecutive cells. We explain the condition of existence of such edge modes below.

 As before, we set $\epsilon=0$ and $t=1$. Labelling the leftmost $B$-type vertex (colored blue in Fig~\ref{binary-ssh} (a)) as $n=1$, and assuming that, the amplitude of the wave function at the leftmost ($B_n$-type) vertex $n=1$ (that has a coordination number equal to four),  viz, $\psi_1$ to be finite, we find that for $E'=0$ in the square-root mapped SSH chain, that is, for $E = \pm 2$ in the parent diamond array, the next amplitude $\psi_2=0$. This is easily verified using Eq,~\eqref{rgeqn}. This immediately relates the amplitudes at all the odd numbered sites on the mapped SSH chain through the relation, 
\begin{equation}
    \psi_{2n+1} = (-1)^n \left [ \frac{\cos 2\theta_1}{\cos 2\theta_2} \right ]^n \psi_1
    \label{converge}
\end{equation}
while, by virtue of $\psi_2=0$ the amplitudes $\psi_{2n} =0$. For a convergent wave function, we must have 
\begin{equation}
    \left |\frac{\cos 2\theta_1}{cos 2\theta_2} \right | < 1
\label{cond}
\end{equation}

%%%%%%%%%%%%%%%%%%%%%%%%%%%%%%%%%%%%%%%%%%%%%%%%%%%%%%%%%%%
\begin{figure*}[ht]
(a)\includegraphics[width=0.7\columnwidth]{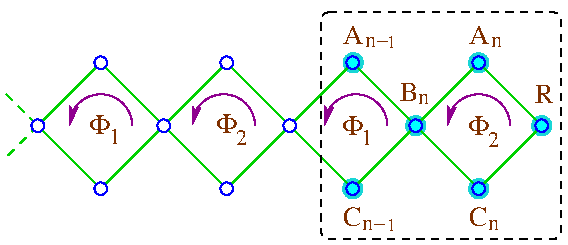}
(b)\includegraphics[width=0.6\columnwidth]{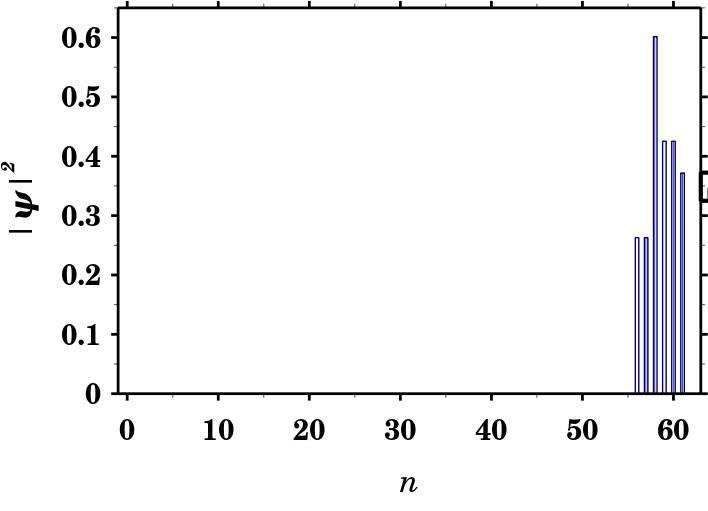}\\
(c)\includegraphics[width=0.7\columnwidth]{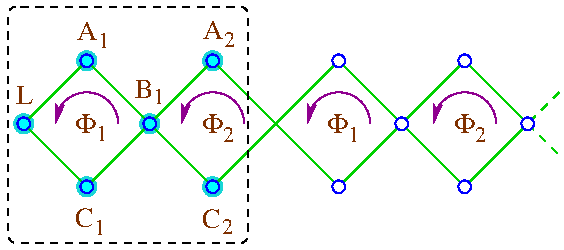}
(d)\includegraphics[width=0.6\columnwidth]{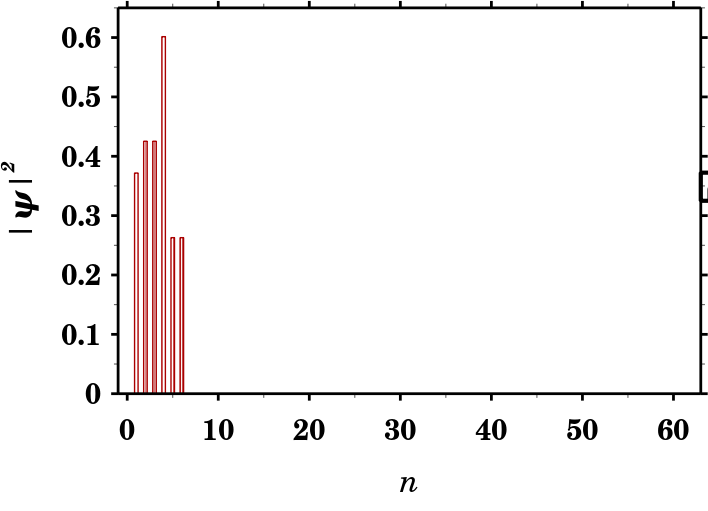}
\caption{(Color online) 
The edge states for a $20$-cell long binary flux staggered DTQN with open boundary conditions. (a,b) $\Phi_1 = \Phi_0/2$, $\Phi_2= \Phi_0$,
and (c,d) $\Phi_1 = \Phi_0$, $\Phi_2 = \Phi_0/2$. The amplitudes are sharply confined at either edge of the network for energy $E = 2.28825$.
The wave function amplitudes are, (a)
$\psi_{A_{n-1}} = 0.262865~e^{-i \pi /4}, \psi_{B_{n}} = 0.601501, \psi_{C_{n-1}} = \psi_{A_{n-1}}^{\ast}, \psi_{A_{n}}= 0.425325~e^{i \pi/2}, \psi_{B_{2}} = -0.371748$, $\psi_{C_{n}}=\psi_{A_{n}}^{\ast}$. 
For (c) we have found a similar configuration, but with different numerical values, as mentioned in the text.
The confinement topography of amplitudes for other edge states with energies $E = -2.28825,\pm 0.87032$
follows similar pattern.}
\label{binary-edge}
\end{figure*}
%===========================================================================
If the condition laid out by Eq.~\eqref{cond} is fulfilled, {\it only then} we have a state localized sharply at site number $n=1$ on the equivalent one dimensional SSH chain. In terms of the parent diamond lattice we should expect that at $E=\pm 2$, the amplitudes will spread over a minimal number of vertices around the extreme left edge. These states should be topologically protected, thanks to the arguments catering a pure,  one dimensional SSH model. 

A similar argument leads to the conclusion that, for a flipped convergence condition, viz, $|\cos \theta_2/\cos \theta_1| < 1$, we shall have an edge state localized around the right-most vertex of a finite diamond array with dangling atoms. However, one must appreciate a very important issue that makes this analytically exact treatment work. That is the assumption that {\it all the sites} ($B_n$) of the renormalized one dimensional {\it effective} SSH chain obtained after decimation, are {\it equivalent} (just like a purely one dimensional SSH lattice where we could set $\epsilon_n=0$ for all $n$). This is precisely what we achieve, when we have a diamond array with two dangling bonds at each end. 

The topologically protected edge states at $E=\pm 2$ are shown in Fig.~\ref{binary-ssh}(c) in red, as the hopping $\tau_1$ is varied, and are seen to lie frozen in the spectral gaps. For this particular figure we have set $\Phi_2=0$, that makes $\tau_2=2$ in the effective (mapped) SSH chain, but the scheme and the result remain similar for any other value of $\Phi_2$ as long as we satisfy the condition Eq.~\eqref{cond}. The robustness of the state against a varied set of values of the hopping $\tau_1$ implicitly implies a continuous distortion in the lattice, and subsequently, in the Hamiltonian. The edge state is thus protected topologically. 

\subsubsection{Chiral edge states in finite DTQN without dangling atoms}

Needless to say that, if we remove the dangling bonds from both the ends, we `perturb' the system, and this perturbation shifts the edge state energy from $E=\pm 2$, and in addition, generates other energies where one can explore different distributions of the edge state amplitudes. This variant of the DTQN, without the dangling bonds, is described in Fig.~\ref{binary-ssh}(b). The corresponding edge states at different energy values show up as $\tau_1$ is varied, keeping $\tau_2=2$, are shown in Fig.~\ref{binary-ssh}(d).

As numerical support to our arguments above, and to unravel chiral edge states, we present some results in Fig.~\ref{binary-edge} where the edge states correspond to FB's. The DTQN chosen doesn't have any dangling edges now.

A finite cluster of atomic sites with non zero amplitudes shows the localization of single particle states {\it at one edge} of the system. It is easy to check that in Fig.~\ref{binary-edge} (c), the cluster of non-zero amplitudes hosts  $\psi_{A_1} = -0.205956+0.372134i$, $\psi_{L} = 0.325257+0.180012i$, $\psi_{C_1} = -\psi_{A_1}^\ast$,  $\psi_{A_2} = -0.0726228-0.252635i$, $\psi_{B_1} = -0.526278-0.291265i$, and $\psi_{C_2} =-i\psi_{A_2}$. Rest of the lattice points experience zero amplitudes.
%%%%%%%%%%%%%%%%%%%%%%%%%%%%%%%%%%%%%%%%%%%%%%%%%%%%%%%%%%%%%%%%%%%%%%%%%%%%
\begin{figure*}[ht]
(a)\includegraphics[width=0.5\columnwidth]{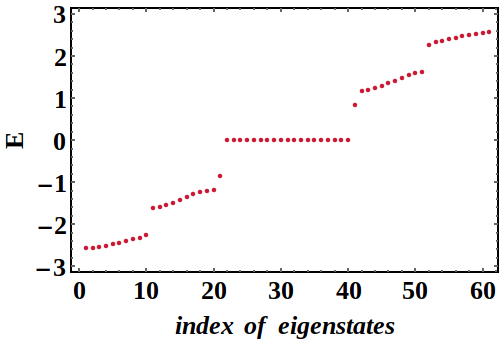}
(b)\includegraphics[width=0.5\columnwidth]{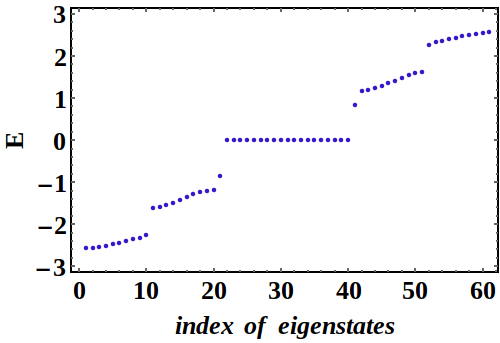}
(c)\includegraphics[width=0.5\columnwidth]{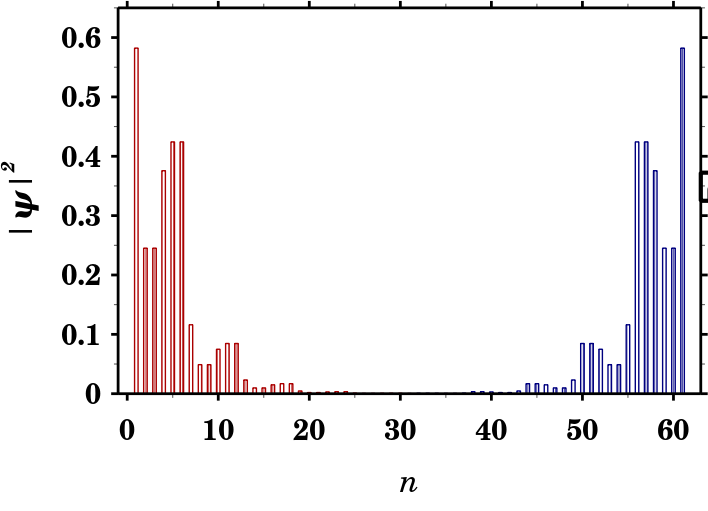}
\caption{(Color online) 
Edge states for a finite diamond chain with binary flux distribution, (a) $\Phi_1 = \Phi_0$, $\Phi_2 = 3\Phi_0/5$ (red) and (b) $\Phi_1 = 3\Phi_0/5$, $\Phi_2 =  \Phi_0$ (blue). The amplitudes exhibit chirality, being trapped over a cluster of a few atomic sites around one edge only for $E = \pm 0.842308$, as shown in the same panel (c).
These in-gap boundary states are $21^{st}$ and $41^{st}$ contributions from the finite assemble of sixty one number of single particle states. }
\label{bi-edge-nonflat}
\end{figure*}
%++++++++++++++++++++++++++++
The FB condition is automatically accomplished by setting either $\Phi_1$ or $\Phi_2$ equal to the half flux quantum. The states reported above are {\it chiral} in the sense that, they are confined to any one edge of the lattice, without having a counterpart sitting at the other end. However, such chiral edge states need not necessarily be associated to non-dispersive bands, as can be seen in Fig.~\ref{bi-edge-nonflat}. In this second example, the span of \textit{confining cluster} becomes larger, encompassing many more lattice points, but localizing eventually. This implies a \textit{spatially delayed} localization, and an extended quantum prison in the spirit of the AB caging effect. The importance of the FB states and the topological edge modes has already been appreciated in literature~\cite{ricardo3}.

\begin{figure*}[ht]
(a)\includegraphics[width=0.99\columnwidth]{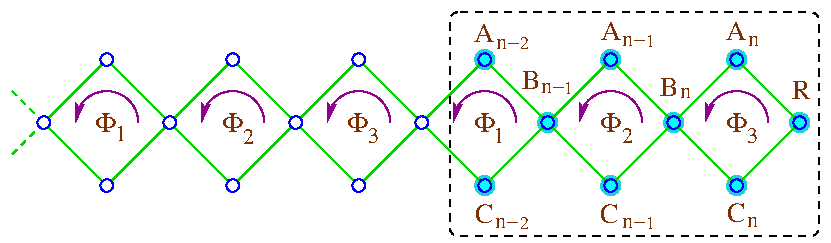}
(b)\includegraphics[width=0.55\columnwidth]{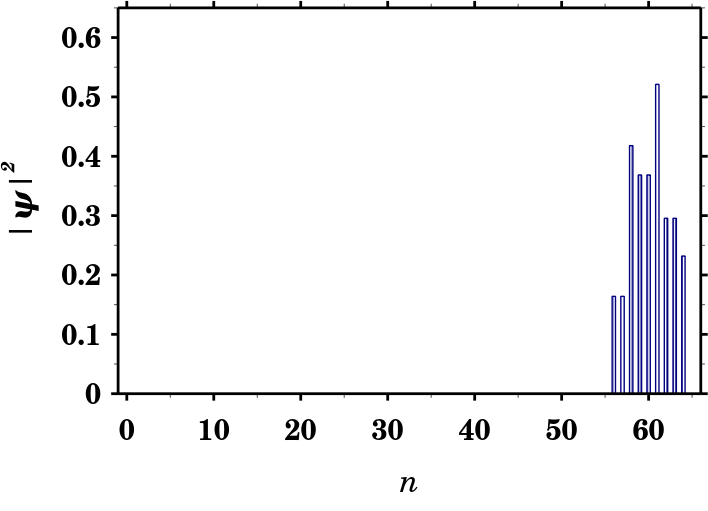}\\
(c)\includegraphics[width=0.99\columnwidth]{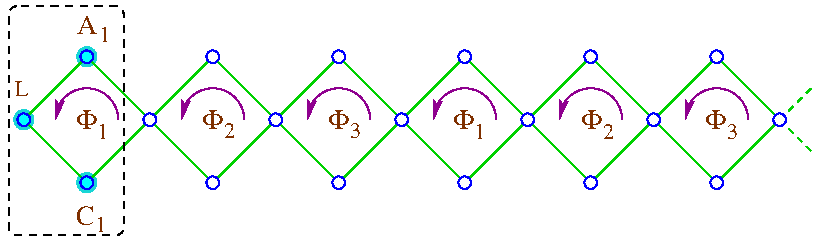}
(d)\includegraphics[width=0.55\columnwidth]{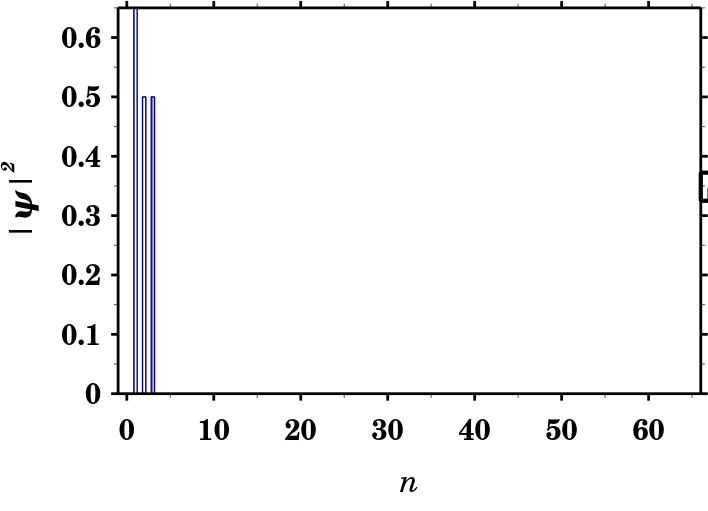}\\
(e)\includegraphics[width=0.99\columnwidth]{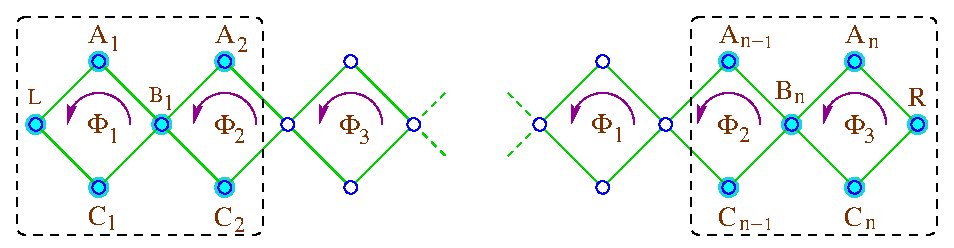}
(f)\includegraphics[width=0.55\columnwidth]{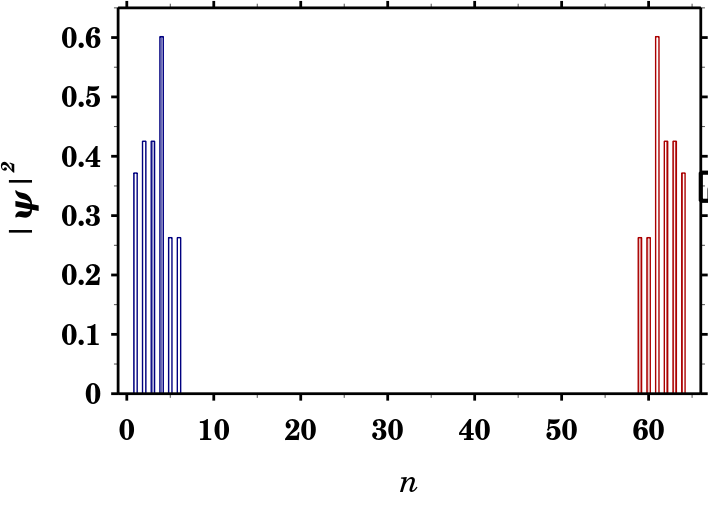}\\
\caption{(Color online) 
Edge states for a diamond array with tertiary flux distribution for a finite size system (a,b,c,d) with $\Phi_1 = 0.5 \Phi_0$, $\Phi_2 = \Phi_3 = \Phi_0$, and
(e,f) $\Phi_2 = 0.5 \Phi_0$, $\Phi_1 = \Phi_3 = \Phi_0$. (a,b) The amplitudes are strictly pinned at one edge (right) of the system for $E = 2.54832$ and for some other energies $E = -2.54832, \pm 1.736, \pm 0.629384$. (c,d) Two chiral edge states with $E=\pm \sqrt{2}$ are confined only at left edge of the system.
(e,f) The amplitudes are strictly pinned at both edges of the system for $E = 2.28825$ and similar distribution for other edge states with energies $E=-2.28825,\pm 0.87032$.
(a) is for $E = 2.54832$ with $\psi_{A_{n-2}} = 0.16399 e^{-i \pi/4}$, $\psi_{B_{n-1}} = 0.417907$, $\psi_{C_{n-2}} = \psi_{A_{n-2}}^{\ast}$, 
$\psi_{A_{n-1}}= 0.368488 e^{i \pi/2}$, 
$\psi_{B_n} = -0.521121$,
$\psi_{C_{n-1}} = 0.368488 e^{-i \pi/2}$,
$\psi_{A_n} =0.295505 e^{-i \pi/2}$,  $\psi_{R} = 0.23194$,
$\psi_{C_n} =0.295505 e^{i \pi/2}$.
(c) is for the energy $E = \sqrt{2}$ with $\psi_{A_1} = -(0.111491+0.487411i)$, 
$\psi_{L} = -(0.598902+0.375921i)$, 
$\psi_{C_1} = -i\psi_{A_1}$.
(e) is for the energy $E = 2.28825$ with $\psi_{A_1} = -(0.406998+0.12351i)$, $\psi_{L} = -0.107951+0.355729i$, $\psi_{C_1} = - \psi_{A_1}$,
 $\psi_{B_1} = 0.174669-0.575582i$, $\psi_{A_2} = 0.23184-0.123889i$, $\psi_{C_2} = -i\psi_{A_2}$, 
 $\psi_{A_{n-1}} = 0.262865 e^{-i \pi/4}$, $\psi_{C_{n-1}} = \psi_{A_{n-1}}^{\ast}$, $\psi_{B_n} = 0.601501$,
 $\psi_{A_n} = 0.425325e^{i \pi/2}$, $\psi_{R} = -0.371748$,
 $\psi_{C_n} = \psi_{A_n}^{\ast}$.
The number of diamond cells is $21$.}
\label{tri-edge}
\end{figure*}

Many more interesting engineering options open up as one heads for increased staggering periods. 
For example, in case of a tertiary staggered flux distribution with $\Phi_1$-$\Phi_2$-$\Phi_3$ repeating periodically (Fig.~\ref{tri-edge}), depending upon which cell traps a half flux quantum, the envelope of the edge modes can exhibit a diverse span of localization. For some combination of flux values, it is localized, say, around the left edge and an inversion in the flux distribution inverts the scenario, again revealing the chiral character of the edge states. Interestingly, if, in a flux period of $\Phi_1$-$\Phi_2$-$\Phi_3$  we select $\Phi_2=\Phi_0/2$, then edge states are symmetrically pinned at both the edges. Thus, by strategically choosing the distribution distribution of magnetic flux, or equivalently, by choosing the angles of axial twist while keeping the external magnetic field constant, both being controlled externally, one can engineer chiral as well as non-chiral edge modes in this quasi-one dimensional quantum network. A selected set of such edge modes and their flux-tunability are displayed in the self explanatory Fig.~\ref{tri-edge}. 

%===========================================================================
%%%%%%%%%%%%%%%%%%%%%%%%%%%%%
\section{Conclusion}
\label{summary}
We have undertaken an in-depth study of a quantum network with diamond shaped cells pierced by a staggered magnetic flux distribution. The state-of-the art photonics has already developed networks trapping synthetic gauge fields. A Floquet engineering of the staggering effect in the distribution of magnetic flux or rather, its {\it synthetic} equivalent may no longer be a remote possibility. The primary interest is to scrutinize the possibility of engineering a topological phase transition by tuning the magnetic flux, an {\it externally controllable} parameter, rather than playing around with the system's intrinsic parameters. The second, and equally important point of interest has been an analysis of the Aharonov-Bohm caging effect along with the occurrence of flat, non-dispersive bands - hallmarks of an extreme localization, that have found immense importance in recent literature. In the case of a binary flux staggered model we use a simple real space decimation technique to map the system onto an effective Su-Schreiffer-Heeger chain and show that, it is indeed possible to monitor the magnetic flux to engineer a topological phase transition. This can happen even in the absence of any quantized topological invariant in the system. The existence of non-chiral and chiral edge modes when the flux pattern is uniform, or show a non-trivial periodicity, is discussed along with the prospect of achieving a comprehensive control over the spatial locations of such edge modes, using a magnetic flux.  The Aharonov-Bohm caging of single particle states, and the consequential extreme localization of the single particle states can be induced by special values of the magnetic flux trapped in the cells of the network. The compact localized states seen in such a phenomenon are nowadays drawing huge attention as potential candidates for transportation of quantum information. We discuss the elementary case of a uniform flux and its variants, and propose a generalization where the flux pattern follows an Aubry-Andr\'{e}-Harper kind of modulation in its commensurate limit. This modulation mimics a continuous axial twist of the entire quantum network immersed in a constant magnetic field, and hence opens up, to our mind,  new and interesting physics related to what we may call, a  {\it twist-induced} topological and localization properties in low dimensional lattice model - an issue that we wish to investigate further and report elsewhere.
%============================================================================

\begin{acknowledgments} 
A. M. acknowledges DST for providing her INSPIRE Fellowship $[IF160437]$. A. M. and A.C. acknowledge stimulating discussions with Sebabrata Mukherjee. Both A. M. and A. N. are grateful to Presidency University for providing the computational facility.
\end{acknowledgments}
\newpage
\appendix
\section{Topological phase transition in the binary flux staggered lattice }
After decimation of the top and bottom vertices $A_n$ and $C_n$ in the diamond cells we arrive at a renormalized version of the array, that effectively resembles an SSH chain, and is described by the difference equation, Eq~\eqref{rgeqn}. This is exactly an SSH chain as we have two flux values $\Phi_1$ and $\Phi_2$ repeating periodically, and have effective nearest neighbor hopping integrals $\tau_1$ and $\tau_2$, explained in the main text.
%%%%%%%%%%%%%%%%%%%%%%%%%%%%%%%%%%%%%%%%%%%%%%
\begin{figure}[ht]
\includegraphics[width=0.95\columnwidth]{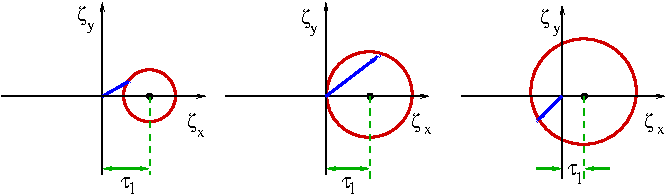}
\caption{(Color online) 
Flow of the winding circle in the $\zeta_x-\zeta_y$ space.}
\label{flow}
\end{figure}
%%%%%%%%%%%%%%%%%%%%%%%%%%%%%%%%%%%%%%%%%%%%%%%
Let us suppose that, we are given this equation to start an analysis on the existence of a topological phase transition. We can design a Hamiltonian describing the unit cell of this effective SSH chain that leads to such an equation. The Hamiltonian is:

\begin{eqnarray}
\mathbf{H}^\ast =  
\left( \begin{array}{cc}
0 & \tau_1 +\tau_2 e^{-ika}\\ 
\tau_1 +\tau_2 e^{ika} & 0
\end{array}
\right)
\label{hamil}
\end{eqnarray}
Here, $\tau_1 = 2 \cos (\pi \Phi_1/\Phi_0)$ and $\tau_2 = 2 \cos (\pi \Phi_2/\Phi_0)$.
The eigenvalues are given by, 
\begin{equation}
    E_{\pm} = \pm  \sqrt{\tau_1^2 + \tau_2^2 + 2 \tau_1 \tau_2 \cos k}
    \label{ssheigen}
\end{equation}
The Bloch eigenvectors corresponding to the eigenvalues can be written in the form,
\begin{eqnarray}
|u_{\pm}(k) \rangle = \frac{1}{\sqrt{2}}
\left( \begin{array}{c} e^{\mp i\phi(k)} \\ 1 
\end{array} \right )
\label{eigvec}
\end{eqnarray}
where, the phase factor $\phi$ is an explicit function of the flux trapped in the cells, and given by,
\begin{equation}
\tan \phi(k) = \frac{\cos (\pi\Phi_2/\Phi_0) \sin ka}{\cos (\pi\Phi_1/\Phi_0)+
\cos (\pi\Phi_2/\Phi_0) \cos ka}
\end{equation}
This formulation is well known for an SSH chain.~\cite{asboth} We exploit this formulation to express our Hamiltonian in a space of {\it reduced degree of freedom} (as a result of the decimation of a subset of vertices) as $H^{\ast} = \vec{\zeta} \cdot \vec{\sigma}$, where, $\vec{\sigma}$ is the vector of the Pauli matrices, and $\vec{\zeta}$ is a three dimensional vector with components, 
\begin{eqnarray}
\zeta_x(k) & = & 2 \cos (\pi \Phi_1/\Phi_0) + 2 \cos (\pi \Phi_2/\Phi_0) \cos ka \nonumber \\
\zeta_y(k) & = & 2 \cos (\pi \Phi_2/\Phi_0) \sin ka \nonumber \\
\zeta_z(k) & = & 0
\label{compo}
\end{eqnarray}

The ``direction'' and the ``norm'' of the vector $\vec\zeta$ contain the information about the eigenstate and its corresponding eigenvector.
It is now straightforward to observe that, the `tip' of the vector $\vec{\zeta}$ traces out a closed loop, which is a circle in this case, in the $\zeta_x-\zeta_y$ plane. The center of the circle being given by the point $[2\cos(\pi\Phi_1/\Phi_0),0]$. The position of the centre is now easily engineered by tuning the external magnetic field. The radius of the circle is $2\cos(\pi\Phi_2/\Phi_0)$, and it is tunable by the flux in an adjacent cell.

The above parametrization and picturization immediately make one conclude that a tuning of the flux values $\Phi_1$ and $\Phi_2$ is equivalent to {\it deforming} the Hamiltonian in a continuous way. Thus, one can drive the system from one insulating phase corresponding to $\Phi_1 < \Phi_2$ to $\Phi_1 > \Phi_2$ by engineering the flux alone. A  topological phase transition is definitely on the cards, and happens exactly at $\Phi_1=\Phi_2$. The circle in the $\zeta_x-\zeta_y$ plane touches the origin as flux trapped in one cell equals the other. This can of course, be attained for a continuous distribution of $\Phi_1$ and $\Phi_2$. The flow of the trajectory in the $\zeta_x-\zeta_y$ space is shown in Fig.~\ref{flow}.

\begin{thebibliography}{0}%
\makeatletter
\providecommand \@ifxundefined [1]{%
 \@ifx{#1\undefined}
}%
\providecommand \@ifnum [1]{%
 \ifnum #1\expandafter \@firstoftwo
 \else \expandafter \@secondoftwo
 \fi
}%
\providecommand \@ifx [1]{%
 \ifx #1\expandafter \@firstoftwo
 \else \expandafter \@secondoftwo
 \fi
}%
\providecommand \natexlab [1]{#1}%
\providecommand \enquote  [1]{``#1''}%
\providecommand \bibnamefont  [1]{#1}%
\providecommand \bibfnamefont [1]{#1}%
\providecommand \citenamefont [1]{#1}%
\providecommand \href@noop [0]{\@secondoftwo}%
\providecommand \href [0]{\begingroup \@sanitize@url \@href}%
\providecommand \@href[1]{\@@startlink{#1}\@@href}%
\providecommand \@@href[1]{\endgroup#1\@@endlink}%
\providecommand \@sanitize@url [0]{\catcode `\\12\catcode `\$12\catcode
  `\&12\catcode `\#12\catcode `\^12\catcode `\_12\catcode `\%12\relax}%
\providecommand \@@startlink[1]{}%
\providecommand \@@endlink[0]{}%
\providecommand \url  [0]{\begingroup\@sanitize@url \@url }%
\providecommand \@url [1]{\endgroup\@href {#1}{\urlprefix }}%
\providecommand \urlprefix  [0]{URL }%
\providecommand \Eprint [0]{\href }%
\providecommand \doibase [0]{http://dx.doi.org/}%
\providecommand \selectlanguage [0]{\@gobble}%
\providecommand \bibinfo  [0]{\@secondoftwo}%
\providecommand \bibfield  [0]{\@secondoftwo}%
\providecommand \translation [1]{[#1]}%
\providecommand \BibitemOpen [0]{}%
\providecommand \bibitemStop [0]{}%
\providecommand \bibitemNoStop [0]{.\EOS\space}%
\providecommand \EOS [0]{\spacefactor3000\relax}%
\providecommand \BibitemShut  [1]{\csname bibitem#1\endcsname}%
\let\auto@bib@innerbib\@empty
%</preamble>
\end{thebibliography}%


\begin{thebibliography}{99}
\bibitem{ssh} W. P. Su, J. R. Schrieffer, and A. J. Heeger, Phys. Rev. Lett. \textbf{42}, 1698 (1979).
\bibitem{heeger} A. J. Heeger, S. Kivelson, J. R. Schrieffer, and W. P. Su, Rev. Mod. Phys. \textbf{60}, 781 (1988).
\bibitem{tknn} D. J. Thouless, M. Kohmoto, M. P. Nitingale, and M. den Nijs, Phys. Rev. Lett. \textbf{49}, 405 (1982).
\bibitem{prange} R. Prange and S. M. Girvin, {\it The Quantum Hall Effect}, 2nd ed. Springer-Verlag, New York, (1990).
\bibitem{wen} X.-G. Wen, Adv. Phys. \textbf{44}, 405 (1995).
\bibitem{folling} S. F\"{o}lling,  S. Trotzky, P. Cheinet, M. Feld, R. Saers, A. Widera, T. M\"{u}ller and I. Bloch, Nature \textbf{448}, 1029 (2007).
\bibitem{sebby} J. Sebby-Strabley, M. Anderlini, P. S. Jessen, and J. V. Porto, Phys. Rev. A \textbf{73}, 033605 (2006).
\bibitem{poli} C. Poli, M. Bellec, U. Kuhl, F. Mortessagne, and H. Schomerus, Nat. Comm. \textbf{6}, 6710 (2015).
\bibitem{liu} C. Liu, W. Gao, B. Yang, and S. Zhang, Phys. Rev. Lett. \textbf{119}, 183901 (2017).

\bibitem{huber} R. S\"{u}sstrunk and S. D. Huber, Science \textbf{349}, 47 (2015).
\bibitem{meier} E. J. Meier, F. Alex An, and B. Gadway, Nat. Comm. \textbf{7}, 13986 (2016).
\bibitem{jotzu} G. Jotzu, M. Messer, R. Desbuquois, M. Lebrat, T. Uehlinger, D. Greif, and T. Esslinger, Nature \textbf{515}, 237 (2014).
\bibitem{leseleuc} S. de L\'{e}s\'{e}leuc, V. Lienhardt, P. Scholl, D. Barredo, S. Weber, N. Lang, H. P. B\"{u}chler, T. Lahange, and A. Browaeys, Science \textbf{365}, 775 (2019).
\bibitem{trimer} V. M. Martinez Alvarez and M. D. Coutinho-Filho, Phys. Rev. A \textbf{99}, 013833 (2019).
\bibitem{miroshnichenko} C. Li and A. E. Miroshnichenko, Physics \textbf{1}, 2 (2019).
\bibitem{seba1} S. Mukherjee and R. R. Thomson, Opt. Lett. \textbf{40}, 5443 (2015).
\bibitem{seba2} S. Mukherjee, M. Di Liberto, P. \"{O}hberg, R. R. Thomson, and N. Goldman, Phys. Rev. Lett. \textbf{121}, 075502 (2018).
\bibitem{alex1} A. Szameit and S. Nolte, J. Phys. B \textbf{43}, 163001 (2010).
\bibitem{rechtsman} M. C. Rechtsman, J. M. Zeuner, Y. Plotnik, Y. Lumer, D. Podolsky, F. Dreisow, S. Nolte, M. Segev and A. Szameit, Nature \textbf{496}, 196 (2013).
\bibitem{alex2} M. Kremer, I. Petrides, E. Meyer, M. Heinrich, O. Zilberberg, and A. Szameit,
Nat. Comm. \textbf{11}, 907 (2020).
\bibitem{jiang} W. Jiang, S. Zhang, Z. Wang, F. Liu, and T. Low, Nano Lett. \textbf{20}, 1959 (2020).
\bibitem{chen} R. Chen and B. Zhen, Phys. Lett. \textbf{381}, 944 (2017).
\bibitem{ankita} A. Bhattacharya and B. Pal, Phys. Rev. B \textbf{100}, 235145 (2019).
\bibitem{nathan} N. Goldman, D. F. Urban, and D. Bercioux, Phys. Rev. A \textbf{83}, 063601 (2011).
\bibitem{ricardo1} L. Madail, S. Flannigan, A. M. Marques, A. J. Daley, and R. G. Dias, Phys. Rev. B \textbf{100}, 125123 (2019).
\bibitem{ricardo2} A. M. Marques and R. G. Dias, Phys. Rev. B \textbf{100}, 041104(R) (2019).
\bibitem{macdonald} A. J. Macdonald, P. C. W. Holdsworth, and R. G. Melko, J. Phys. Condens. Matt. \textbf{23}, 164208 (2011).
\bibitem{fallani} L. Fallani, C. Fort, J. E. Lye and M. Inguscio, Opt. Exp. \textbf{13}, 4303 (1995).
\bibitem{shreekantha} S. Sil, S. K. Maiti, and A. Chakrabarti, Phys. Rev. B \textbf{79}, 193309 (2009).
\bibitem{aharony1} A. Aharony, O. Entin-Wohlman, Y. Tokura, and S. Katsumoto, Phys. Rev. B \textbf{78}, 125328 (2008).
\bibitem{aharony2} A. Aharony, O. Entin-Wohlman, Y. Tokura, and S. Katsumoto, Physica E \textbf{42}, 629 (2010).
\bibitem{amrita} A. Mukherjee, R. A. R\"{o}mer, and A. Chakrabarti, Phys. Rev. B \textbf{100}, 161108(R) (2019).
%%%%%%%%%  AB Caging and Flat Bands %%%%%%%%%%%%%%%
\bibitem{julien1} J. Vidal, R. Mosseri, and B. Do\c{u}cot, Phys. Rev. Lett. \textbf{81}, 5888 (1998).
\bibitem{julien2} J. Vidal, P. Butaud, B. Do\c{u}ot, and R. Mosseri, Phys. Rev. B \textbf{64}, 155306 (2001).
\bibitem{leykam1} D. Leykam, S. Flach, O. Bahat-Treidel, and A. S. Desyatnikov, Phys. Rev. B \textbf{88}, 224203 (2013).
\bibitem{leykam2} D. Leykam, A. Andreanov, and S. Flach, Adv. Phys. X \textbf{3}, 1473052 (2018).
\bibitem{leykam3} D. Leykam and S. Flach, APL Photonics \textbf{3}, 070901 (2018).
\bibitem{flach1} W. Maimaiti, A. Andreanov, H. C. Park, O. Gendelman, and S. Flach, Phys. Rev. B \textbf{95}, 115135 (2017).
\bibitem{flach2} A. Ramachandran, A. Andreanov, and S. Flach, Phys. Rev. B \textbf{96}, 161104(R) (2017).
\bibitem{maimaiti} W. Maimaiti, S. Flach, and A. Andreanov, Phys. Rev. B \textbf{99}, 125129 (2019).
\bibitem{ajith} N. Myoung, H. C. Park, A. Ramachandran, E. Lidorikis, and J-W. Ryu, Sci. Rep. \textbf{9}, 2862 (2019).
\bibitem{travkin} E. Travkin, F. Diebel, and C. Denz, Appl. Phys. Lett. \textbf{111}, 011104 (2017).
\bibitem{xia} S. Xia, C. Danieli, W. Yan, D. Li, S. Xia, J. Ma, H. Lu, D. Song, L. Tang, S. Flach, and Z. Chen, arXiv.: 1912.09703 (2019).
\bibitem{gligoric} G. Gligori\'{c}, P. P. Beli\u{c}ev, D. Leykam, and A. Maluckov, Phys. Rev. A \textbf{99}, 013826 (2019).
\bibitem{bergman} D. L. Bergman, C. Wu, and L. Balents, Phys. Rev. B \textbf{78}, 125104 (2008).
\bibitem{rhim} J.-W. Rhim and B.-J. Yang, Phys. Rev. B \textbf{99}, 045107 (2019).
\bibitem{rodrigo} R. A. Vicencio, C. Cantillano, L. Morales-Inostroza, B. Real, C. Mej\'{i}a-Cort\'{e}s, S. Weimann, A. Szameit, and M. I. Molina, Phys. Rev. Lett. \textbf{114}, 245503 (2015).
\bibitem{rontgen} M. R\"{o}ntgen, C. V. Morfonios, L. Brouzos, F. K. Diakonos, and P. Schmelcher, Phys. Rev. Lett. \textbf{123}, 080504 (2019).
%%%%%%%%%%%%%%%%%%%%%%%%%%%%%%%%%%%%%%%%%%%%%%%%%%%
\bibitem{zak} J. Zak, Phys. Rev. Lett., \textbf{62}, 2747 (1989).
\bibitem{berry} M. V. Berry, Proc. R. Soc. Lond. A. \textbf{392}, 45 (1984).
\bibitem{atala} M. Atala, M. Aidelsburger, J. T. Barreiro, D. Abanin, T. Kitagawa, E. Demler, and I. Bloch, Nature Phys. \textbf{9}, 795 (2013).
\bibitem{fukui} T. Fukui and Y. Hatsugai, and H. Suzuki, J. Phys. Soc. Jpn. \textbf{74}, 1674 (2005).
\bibitem{marzari} N. Marzari and D. Vanderbilt, Phys. Rev. B \textbf{56}, 12847 (1997).
\bibitem{overarxiv} M. B. de Paz, C. Devescovi, G. Giedke, J. J. Saenz, M. G. Vergniory, B. Bradlyn, D. Bercioux, and A. Garc\'{i}a-Etxarri, Adv. Quantum Technol. \textbf{3}, 1900117 (2020).
\bibitem{arkinstall} J. Arkinstall, M. H. Teimourpour, L. Feng, R. El-Ganainy, and H. Schomerus, Phys. Rev. B \textbf{95}, 165109 (2017).
\bibitem{aubry} S. Aubry and G. Andr\'{e}, Ann. Isr. Phys. Soc. \textbf{3}, 133 (1980).
\bibitem{harper} P. G. Harper, Proc. Phys. Soc. A \textbf{68}, 874 (1955).
\bibitem{thouless} D. J. Thouless, Phys. Rev. Lett. \textbf{61}, 2141 (1988).
\bibitem{sankar1} S. Das Sarma, S. He, and X. C. Xie, Phys. Rev. Lett. \textbf{61}, 2144 (1988).
\bibitem{sankar2} S. Das Sarma, S. He, and X. C. Xie, Phys. Rev. B \textbf{41}, 5544 (1990).
\bibitem{sankar3} J. Biddle and S. Das Sarma, Phys. Rev. Lett. \textbf{104}, 070601 (2010).
\bibitem{sankar4} S. Ganeshan, K. Sun, and S. Das Sarma, Phys. Rev. Lett. \textbf{110}, 180403 (2013).
\bibitem{asboth} J. K. Asb\'{o}th, L. Oroszl\'{a}ny, and A. P\'{a}lya, Springer \textbf{99} (2016).
\bibitem{ricardo3} G. Pelegr\'{i}, A. M. Marques, R. G. Dias, A. J. Daley, J. Mompart, and V. Ahufinger,
Phys. Rev. A \textbf{99}, 023613 (2019).
%-------------------------------------------------------%



\end{thebibliography}

\end{document}